\documentclass[a4paper,11pt]{article}
\pdfoutput=1 

\usepackage{jinstpub} 
\usepackage{xcolor} 
\usepackage{fancyvrb}
\usepackage{lineno}
\usepackage[font=small,labelfont=bf]{caption}
\usepackage{gensymb} 
\usepackage{arydshln} 
\usepackage{graphicx} 
\usepackage{subcaption} 
\usepackage{mwe} 
\usepackage{tablefootnote}

\title{\boldmath Sensitivity estimates for diffuse, point-like, and extended neutrino sources with KM3NeT/ARCA}

\author[a]{B. Caiffi,}
\author[b,c]{A. Garcia Soto,}
\author[b,c]{A. Heijboer,}
\author[a]{V. Kulikovskiy,}
\author[b,c,1]{R.S. Muller\note{Corresponding author.}}
\author[a,d]{and M. Sanguineti}

\affiliation[a]{INFN, \\ Sezione di Genova,\\ Via Dodecaneso 33, 16146 Genova, Italy}
\affiliation[b]{Nikhef,\\Science Park 105, 1098 XG Amsterdam, The Netherlands}
\affiliation[c]{University of Amsterdam,\\Science Park 904, 1098 XH Amsterdam, The Netherlands}
\affiliation[d]{Dipartimento di Fisica dell'Universit\`a,\\ Via Dodecaneso 33, 16146 Genova, Italy}

\emailAdd{rasam@nikhef.nl}

\abstract{The identification of cosmic objects emitting high energy neutrinos could provide new insights about the Universe and its active sources. The existence of these cosmic neutrinos has been proven by the IceCube Collaboration, but the big question of which sources these neutrinos originate from, remains unanswered. The KM3NeT detector for Astroparticle Research with Cosmics in the Abyss (ARCA), with a cubic kilometer instrumented volume, is currently being built in the Mediterranean Sea. It will excel at identifying cosmic neutrino sources due to its unprecedented angular resolution for muon neutrinos (< 0.2 degree for $E > 10$ TeV events). KM3NeT has a view of the sky complementary to IceCube, and is sensitive to neutrinos across a wide range of energies. In order to identify the signature of cosmic neutrino sources in the background of atmospheric neutrinos and muons, statistical methods are being developed and tested with Monte-Carlo pseudo-experiments. This contribution presents the most recent sensitivity estimates for diffuse, point-like, and extended neutrino sources with KM3NeT/ARCA.}


\collaboration[c]{on behalf of the KM3NeT Collaboration}

\proceeding{Very Large Volume Neutrino Telescope Workshop (VLVnT)\\
  21 May 2021\\
  Valencia}

\begin{document}
\maketitle
\flushbottom
\VerbatimFootnotes

\section{Introduction}
\label{sec:intro}

In 2013, the IceCube Neutrino Observatory confirmed the existence of a high-energy astrophysical neutrino flux~\cite{ICC_2013}. However, the origin of most of these neutrinos remains unknown. The KM3NeT/ARCA detector~\cite{LOI} that is currently under construction at the bottom of the Mediterranean Sea, will consist of 2 building blocks, each one comprising 115 lines with 18 digital optical modules per line. It will have an excellent pointing resolution ($<0.2^\circ$ for $E>10$ TeV muon neutrinos), and will be sensitive in a large energy range (GeV - PeV). ARCA sky coverage for upgoing neutrinos will be complementary to the IceCube detector and includes the Galactic Centre.

In order to identify a cosmic neutrino signal on top of the atmospheric background of muons and neutrinos, statistical methods are being developed based on Monte Carlo pseudo experiments. Compared to previously published work~\cite{Agatapaper}, the new methods presented here are the start of a bigger software framework for all future high energy astrophysical analyses. 
After applying the proper cuts to increase the signal to background ratio, the detector response functions are determined. These distributions, like the so called effective area, the energy resolution and angular resolution are converted to probability density functions and are used as an input for the unbinned likelihood analyses to calculate the expected sensitivity of KM3NeT/ARCA to diffuse, point source and extended sources in our universe. The methods and results will be presented in this contribution.

\section{Simulation and event selection}
\label{sec:sim_and_select}
\paragraph{Simulations}
The analyses is based on GENHEN (version v7r6) \cite{genhen} Monte Carlo data simulations of $10^2$--$10^8$ GeV  neutrino interactions for the KM3NeT/ARCA detector. 
Neutrinos are weighted using either cosmic or atmospheric flux models. The latter being divided into a Honda(2006) conventional flux component ~\cite{Honda2006_conventional} and a prompt component parametrisation~\cite{ERS_prompt}, both with Gaidder-H3a knee correction ~\cite{Gaidder-H3a_kneecorrection} applied. The $\nu_{\tau}$ channel is currently not taken into account. Atmospheric muons are simulated by the Mupage package (version v3r4he) \cite{mupage}. Events are subjected to a full simulation of light generation and to the response of the detector, and are processed through the same trigger and track reconstruction algorithms that are foreseen for data.


\paragraph{Event selection}
The analyses uses well reconstructed tracks with a horizontal or upward going ($\theta < 100 \degree$) reconstructed direction. 
On the events that are not selected, a boosted decision tree (BDT) is applied, which is trained using 20 variables related to the track reconstruction (likelihood, angular error, etc.). 
Because the background contribution is larger for horizontal tracks, two different values are chosen to apply the cut for upgoing ($\mathrm{BDT} > 0$) and horizontal tracks ($\mathrm{BDT} > 0.7$).

\paragraph{Performance of the event selection}

\begin{figure}[h]
\centering
\includegraphics[width=0.40\textwidth]{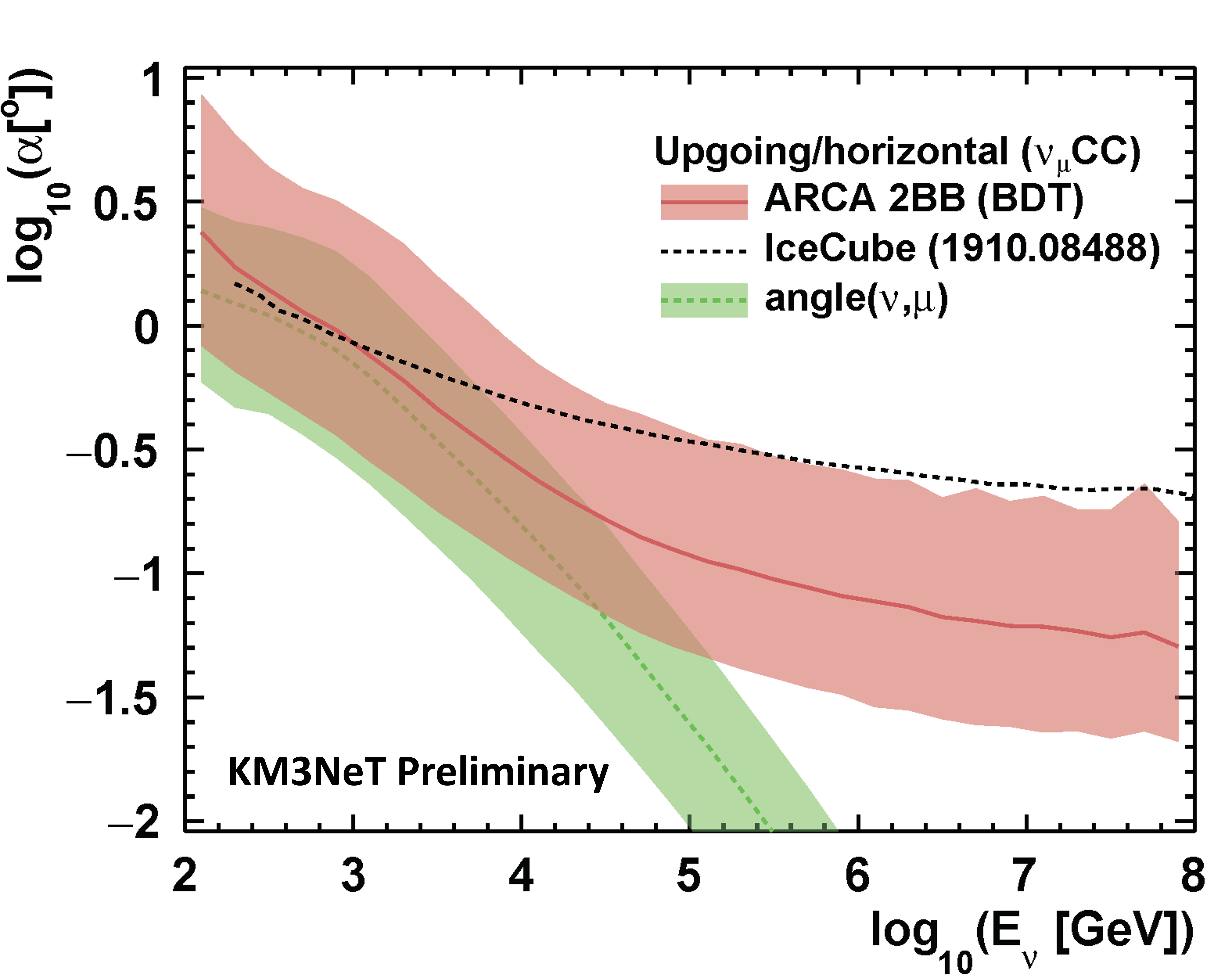}
\includegraphics[width=0.45\textwidth]{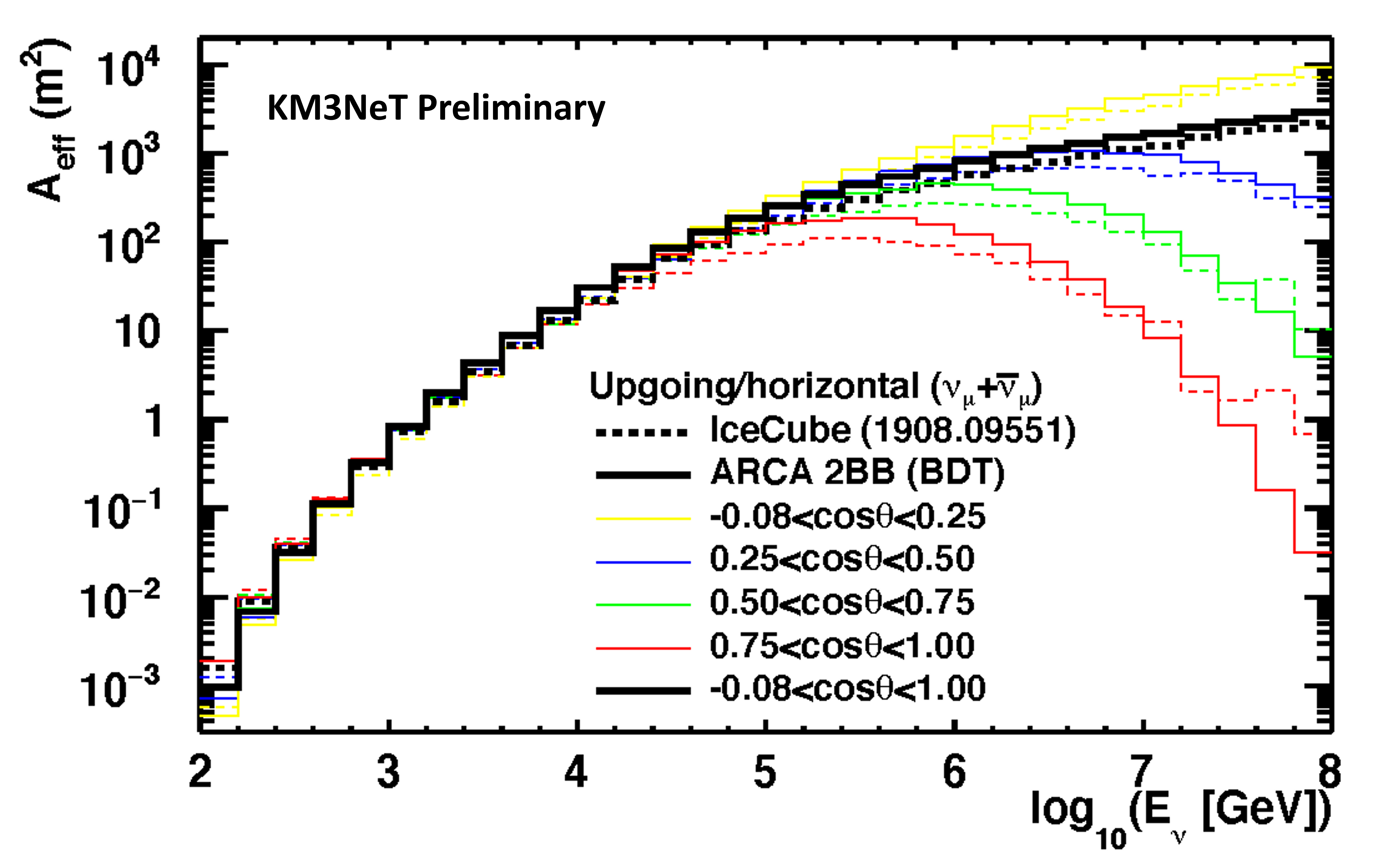}
\caption[]{Left: Angular resolutions as function of the neutrino energy for $\nu_\mu$ CC events. IceCube resolution was extracted from the point source analyses~\cite{ares}. Right: Effective area as function of the neutrino energy for the selected event sample. IceCube effective area was extracted from the diffuse analyses~\cite{aeff}.} 
\label{fig:BDT2}
\end{figure}

The overall performance of the selection as described in this section is presented in Table \ref{tab:final} and Figure \ref{fig:BDT2}. Table \ref{tab:final} shows the number of events with upgoing and horizontal tracks, and those events that fulfil the signal-like criteria, or survive the BDT cut. The final sample contains both these type of events.
The neutrino purity of the final sample, defined as the ratio between the number of selected neutrino events and number of selected events, is 99.4\%.
The signal efficiency, defines as the ratio between number of selected signal events and the number of signal events with upgoing or horizontal tracks,
is computed for each neutrino component independently as a function of the neutrino energy. Overall the efficiency for well reconstructed tracks is above 90\% and for events with muons in the detector above 80\%. One important feature of this selection is that the efficiency is almost identical for atmospheric and cosmic fluxes with different shapes since it is barely energy dependent. Hence, the BDT selection will not be biased towards a certain cosmic flux assumption.

\begin{table}[h]
\centering
\begin{tabular}{l|r:r:r|r:r:r}
      &  \multicolumn{3}{c}{Zenith cut} & \multicolumn{3}{c}{Total}\\
      &  all & +w/muon & +$\alpha<10\degree$ & all & +w/muon & +$\alpha<10\degree$\\
\hline
muons &  208685.6 & & & 249.5 & &  \\
\hline
conv. $\nu$ & 54012.4 & 49937.4 & 43480.8 & 42744.4 & 42687.7 & 41454.4\\
\hline
prompt $\nu$ &  225.3 & 147.9 & 129.3 & 128.5 & 127.6 & 124.4 \\
\hline
cosmic $\nu$ &  158.7 & 113.5 & 101.2 & 99.4 & 98.6 & 97.0\\
\end{tabular}
\caption{Number of expected events for one year of operation with one building block after selecting events with horizontal or upgoing tracks ('Zenith cut' column) and passing the overall selection ('Total' column). +w/muon = Highest $E$ muon in the event must produce at least one hit in two different digital optical modules. $\alpha<10\degree$ = Angle between the reconstructed track and the highest $E$ muon in the event $< 10\degree$.}
\label{tab:final}
\end{table}

\section{Likelihood formalism}
From the selected event sample, detector response functions are created in the form of histogrammed functions that describe the acceptance, the angular resolution and the performance of the energy estimate, and are used as an input for the likelihood analyses. From these distributions, so called pseudo-experiments are generated, and the extended maximum for the hypotheses with and without a cosmic neutrino source are calculated to form the likelihood ratio as the optimal observable to test for the presence of a source. The detector point spread function (PSF) is smeared to account for the source extension.

The global likelihood maximised is given by ${\cal L } = \sum_{i \in \rm events} \sum_{k \in \rm components} \log \bigl[ \frac{d N_{i,k} }{d \Omega d \log(E_{\rm rec})} \bigr] - N^{\rm tot}, $
where the concept of \emph{component} has been introduced, to denote a physical contribution to the event rate (such as background, a diffuse flux, or a point-source flux of neutrinos). $N^{\rm tot}$ denotes the total number of expected events in all flavors, and all channels. The quantity in square brackets is the likelihood of each event $i$ for the component in question. For atmospheric neutrino background, it is taken directly from a distribution of $\theta$ and $E_{\rm rec}$ obtained from simulations of the backgrounds, for the channel corresponding to event $i$. For a point-source component, this can be written as:
\begin{equation}
\label{bigone}
    \frac{d N_i }{d \Omega d \log(E_{\rm rec})} = \sum_{f \in \rm flavors} \int dE \frac{dP_{fc}(E,\theta,\alpha)}{d\Omega d \log(E_{\rm rec})}
    A_{f,c}^{\rm eff} (E,\theta) \Phi_f( E ).
\end{equation}

\section{Results and conclusion}
\label{sec:res_and_concl}

\begin{figure*}[t]
    \centering
    \begin{subfigure}[b]{\textwidth}
        \centering
        \includegraphics[width=0.40\textwidth]{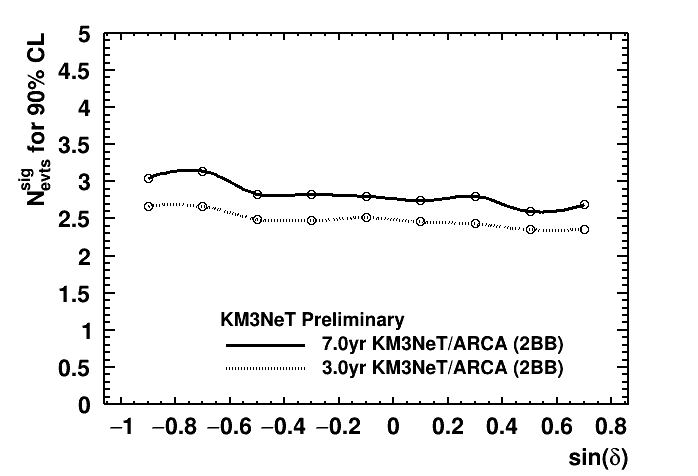}
        \includegraphics[width=0.40\textwidth]{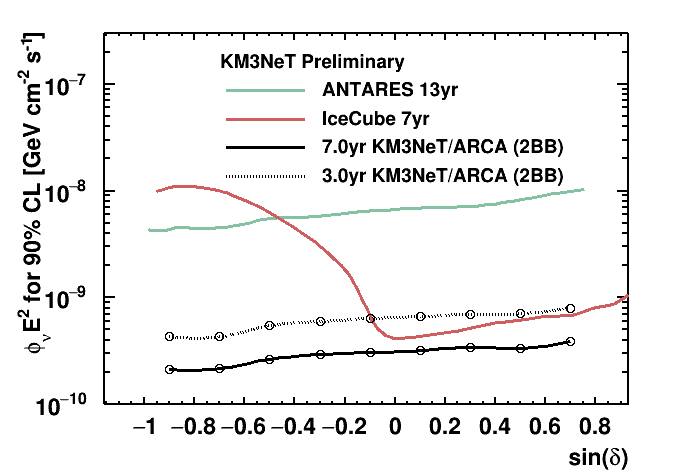}
    \end{subfigure}
    \hfill
    \begin{subfigure}[b]{\textwidth}   
        \centering 
        \includegraphics[width=0.40\textwidth]{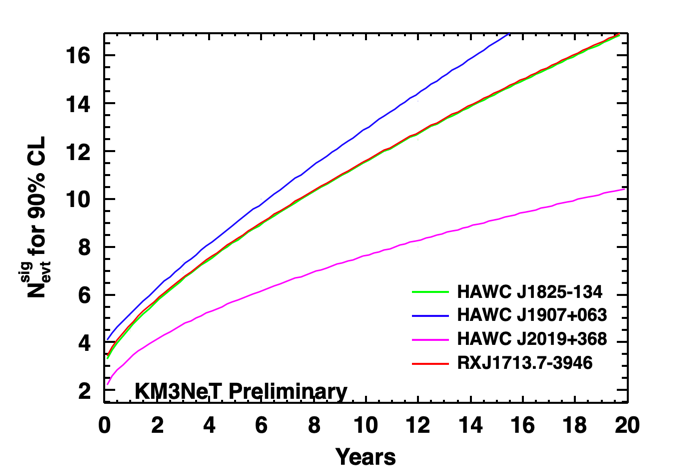}
        \includegraphics[width=0.40\textwidth]{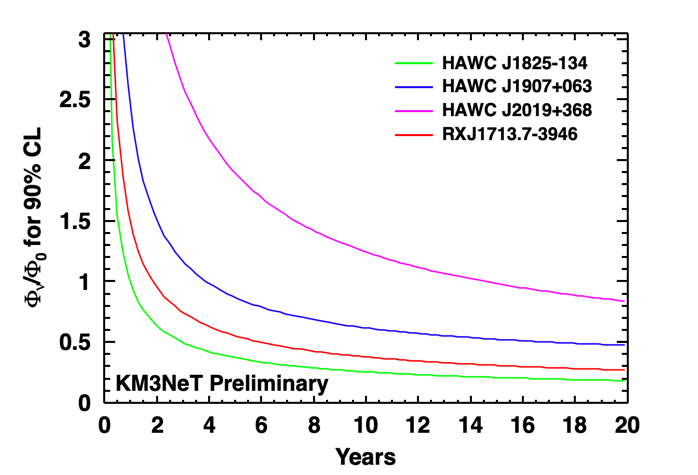}
    \end{subfigure}
    \caption[]
    {\small Number of events (top left) and flux normalisation (top right) for 90\% CL limits to detect a $E^{-2}$ point source with KM3NeT/ARCA in comparison with ANTARES 13 years \cite{ICRC_ANT13}, and IceCube 7 years~\cite{ICC_CL_line_2017}. Bottom left: Average upper limit at 90\% CL. Bottom right: Sensitivity at 90\% CL.} 
    \label{fig:RESULTS}
\end{figure*}

\paragraph{$E^{-2}$ point source analyses}
The sensitivity of KM3NeT to an $E^{-2}$ flux point source at different sky positions is analysed. 
90\% CL exclusion limits are determined based on 40.000 (4000) pseudo experiments for the H0 (H1) distributions. The flux normalisation parameters ($N_{\rm bkg}$, $N_{\rm sig}$) are fitted, whereas the cosmic spectral index ($\gamma = 2$) is kept fixed. In Figure \ref{fig:RESULTS}, the results are shown in comparison to similar studies for 13 years of data taking with ANTARES ~\cite{ICRC_ANT13}, and 7 years for IceCube~\cite{ICC_CL_line_2017}.


\paragraph{Extended source analyses}
The sensitivity of KM3NeT 
to the sources in Table ~\ref{tab:source_data} is estimated.
The prescriptions in~\cite{Vissani} is implemented to derive the expected neutrino flux, as shown in Figure \ref{fig:EXTENDED1}.

   
   

\noindent
   \begin{minipage}[b]{0.5\textwidth}
   \begin{sloppy}
   \raggedleft
    \begin{tabular}{c|c|c}
        Source & Decl, RA & ext \\ 
         & [deg] &  [deg] \\          
        \hline
        RXJ 1713.7-3946 & -39.77, 258.8 & 0.60 (disk) \\
        \hline
        HAWC J1825-134 & -13.37, 276.4 & 0.53 (Gauss) \\
        \hline
        HAWC J1907+063 & 6.32, 286.91 & 0.67 (Gauss) \\
        HAWC J2019+368 & 36.76, 304.92 & 0.356 (Gauss) \\
        \hline
    \end{tabular}
      \captionof{table}{Used source parameters and morphology. (Refs top to bottom: ~\cite{Aharonian_2006}, ~\cite{Abeysekara_2020}, ~\cite{Abeysekara_2020}, ~\cite{hawccollaboration2021spectrum}) }
      \label{tab:source_data}
    \end{sloppy}
    \end{minipage}%
      \hfill
  \begin{minipage}[b]{0.40\textwidth}
   
    \includegraphics[width=\columnwidth]{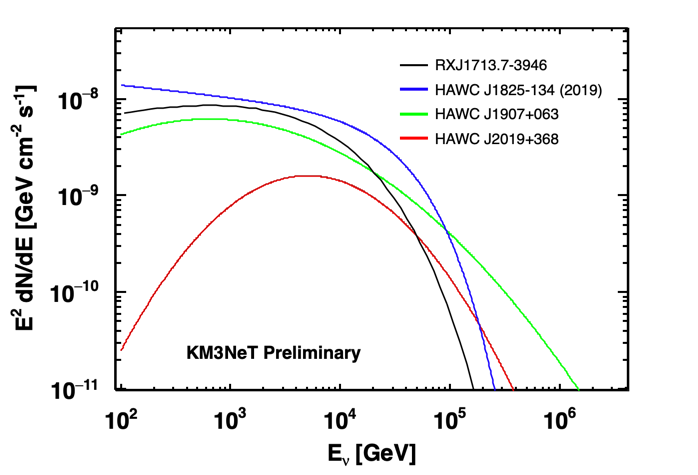}
   
    \captionof{figure}{Expected neutrino fluxes of the sources (assume 100\% hadronic scenario).}
    \label{fig:EXTENDED1}
  \end{minipage}%

The results 
are summarised in Figure \ref{fig:RESULTS}. This study demonstrates the capability of the KM3NeT detector to achieve a 90\% CL sensitivity for 3 of the 4 considered sources in less than 4 years; for the most promising source (HAWC J1825-134) the sensitivity is achieved approximately in 1 year.

\begin{figure}[h!]
\centering
\includegraphics[width=0.4\textwidth]{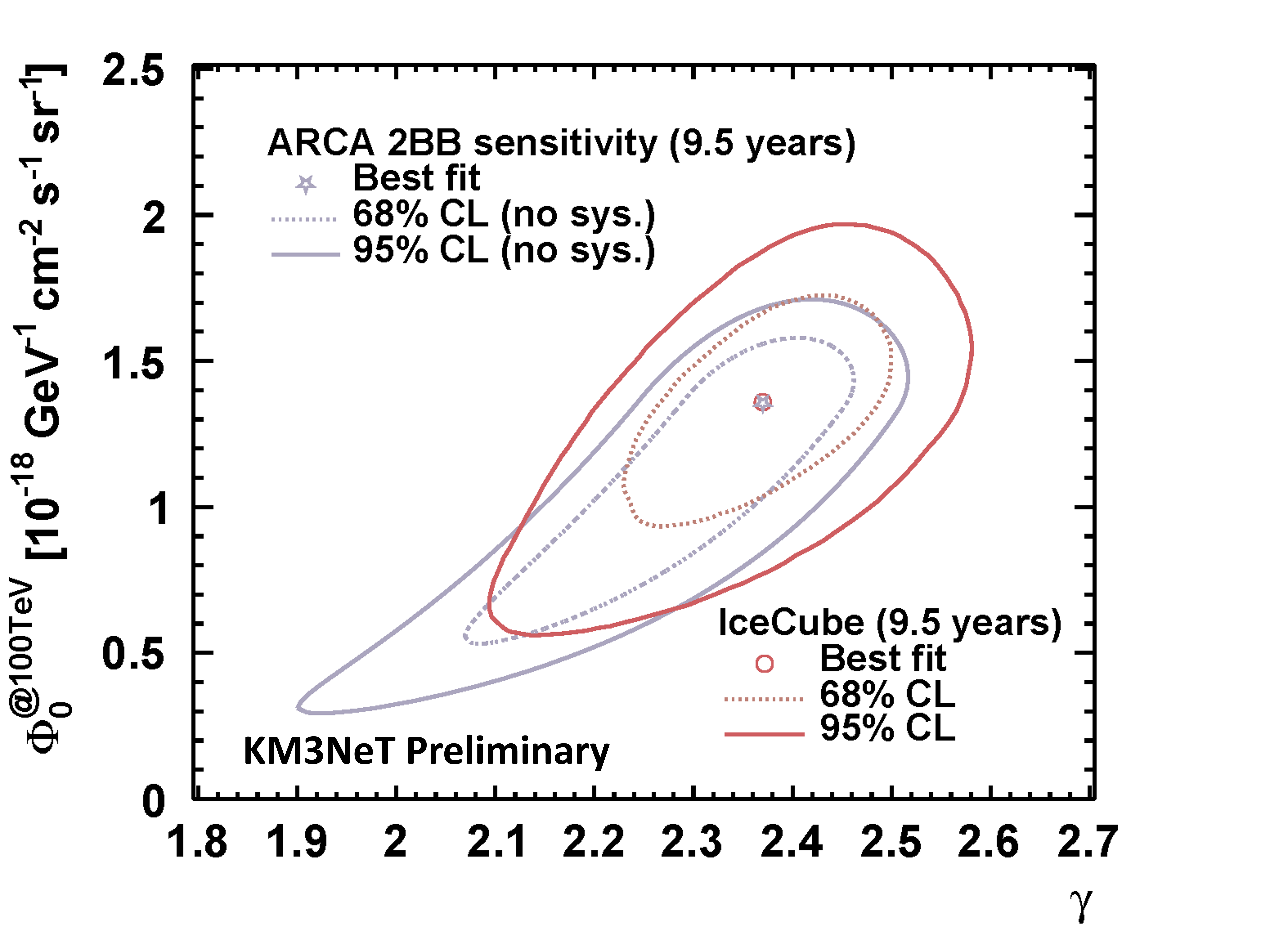}
\includegraphics[width=0.4\textwidth]{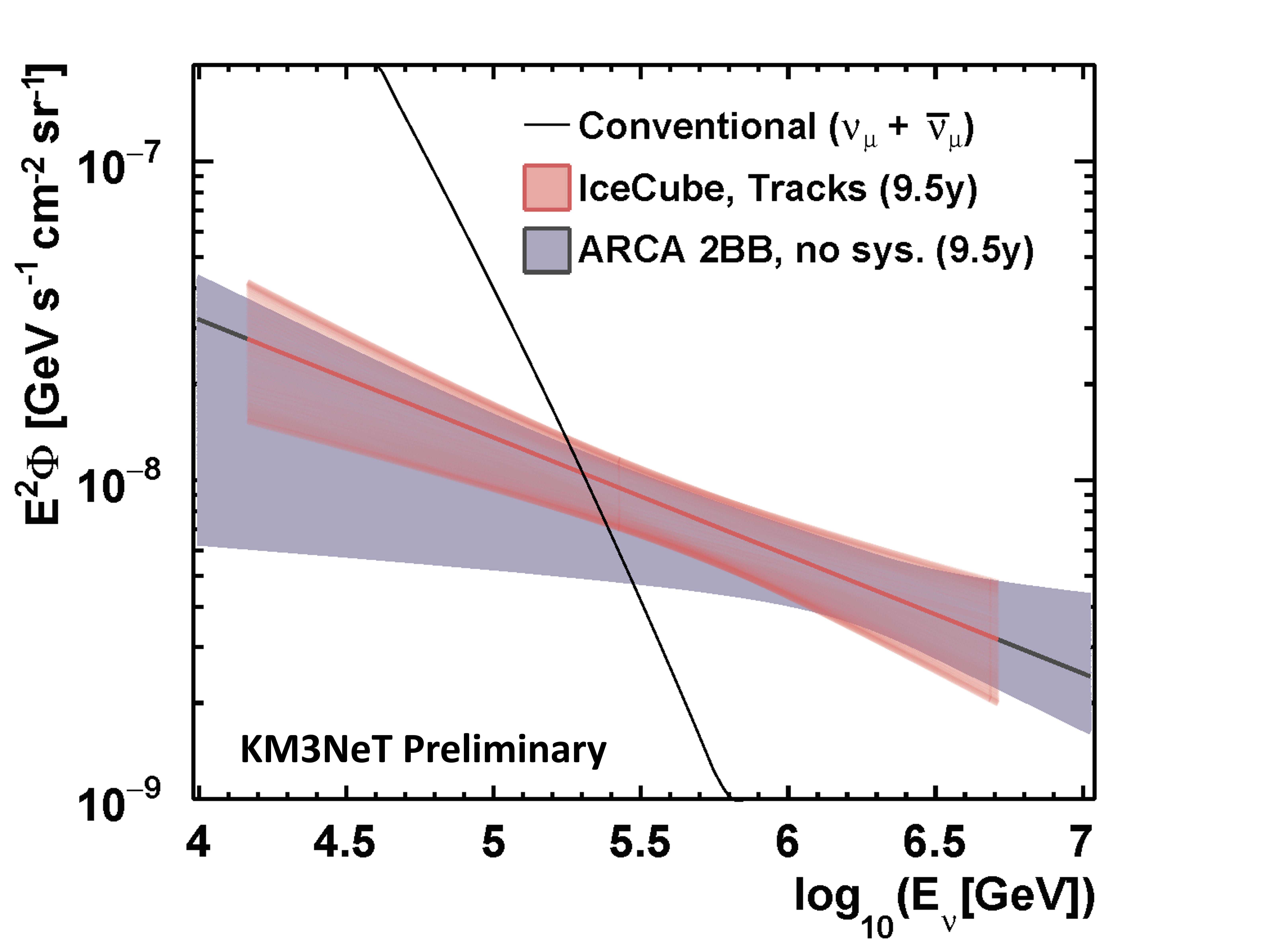}
\caption[]{Left: 2D confidence regions of the astrophysical parameters. Right: Neutrino spectra for the diffuse flux. In gray: 68\% confidence interval for 9.5 years of KM3NeT/ARCA (no systematics). In red: best fit from IceCubes analyses~\cite{icecube_diffuse}. The black line: the conventional atmospheric neutrino flux prediction~\cite{Honda2006_conventional}.}
\label{fig:DIFFUSE}
\end{figure}

\paragraph{Diffuse flux analyses}
To characterise the astrophysical neutrino flux measured by IceCube's diffuse analyses~\cite{icecube_diffuse} 
our Asimov dataset~\cite{asimov} is generated assuming a nominal conventional flux, zero prompt flux and single power law spectra for the cosmic flux ($\Phi = 0.963\cdot10^{-6} E^{-2.37} ~[\rm GeV ^{-1} \rm cm ^{-2} \rm s ^{-1} \rm sr ^{-1}]$). $\Phi_{\rm conv}$,~$\Phi_{\rm prompt}$ and~$\Phi_{0}$, and the spectral slope of the cosmic flux ($\gamma$) are fitted. 
In Figure \ref{fig:DIFFUSE}, the expected sensitivity of KM3NeT/ARCA is compared with similar IceCube result. The narrower contours of this analyses are mainly due to the lack of systematic uncertainties. The elongated shape is due to a different modelling of the prompt component in the IceCube analyses~\cite{MCEQ} and 
~\cite{ERS_prompt}.

\paragraph{In conclusion} Our new methods for high energy sensitivity studies for diffuse, point-like and extended neutrino sources with KM3NeT/ARCA is working and providing convincing first results.




\end{document}